\documentclass[epj]{svjour}
\usepackage{graphics}
\usepackage{amsmath}
\usepackage{amssymb}

\begin{document}

\def\bb    #1{\hbox{\boldmath${#1}$}}
\def\square {\mathchoice\sqr34\sqr34\sqr{2.1} 3\sqr{1.5} 3}

%
\title{QED\,Meson\,Description\,of\,the\,X17\,and\,Other\,Anomalous\,Particles}
\author{Cheuk-Yin Wong\inst{} 
}                     
%
%
\institute{Physics Division, Oak Ridge National Laboratory\thanks{
This manuscript has been authored in part by UT-Battelle, LLC, under
contract DE-AC05-00OR22725 with the US Department of Energy (DOE). The
US government retains and the publisher, by accepting the article for
publication, acknowledges that the US government retains a
nonexclusive, paid-up, irrevocable, worldwide license to publish or
reproduce the published form of this manuscript, or allow others to do
so, for US government purposes. DOE will provide public access to
these results of federally sponsored research in accordance with the
DOE Public Access Plan
(http://energy.gov/downloads/doe-public-access-plan), Oak Ridge,
Tennessee 37831, USA }, ~\,Oak Ridge, Tennessee 37831, USA }
\date{Received: date / Revised version: date}
%
\abstract{
The X17 particle, the E38 particle, and the anomalous soft photons are
anomalous particles because they do not appear to belong to any known
Standard Model families.  We propose a QED meson description of the
anomalous particles as composite systems of a light quark and a light
antiquark bound and confined by the compact QED interaction, by
combining Polyakov's transverse confinement of opposite electric
charges in compact QED in (2+1)D and Schwinger's longitudinal
confinement for massless opposite electric charges in QED in (1+1)D.
With predicted QED meson masses close to the observed X17 and E38
masses, QED mesons may be good candidates for the description of the
anomalous particles.
} 
\maketitle
\section{Introduction}
\label{intro}

The observed X17 particle \cite{Kra16,Kra21}, the E38 particle
\cite{Abr12,Abr19}, and the anomalous soft photons
\cite{Chl84,Bot91,Bel02,Per09,DEL10} are anomalous particles because
their masses of many tens of MeV do not lie in the mass region of any
known family of particles of the Standard Model.  Many different
interpretations have been presented and their theoretical implications
discussed \cite{X1722}.  We focus our attention on the description of
the X17 particle and other anomalous particles as composite particles
of a light quark and a light antiquark bound and confined by their
mutual QED interaction \cite{Won10,Won20,Won20a,Won21,Kos21}.  We
shall call such a composite particle a QED meson (compactly written as
a ``{\it qedmeson}''), in analogy with the QCD meson.

\hspace*{-0.2cm}
Previously, in many exclusive experiments in high-energy hadron-hadron
collisions and $e^+ e^-$ annihilations, it has been consistently
observed that whenever hadrons are produced, anomalous soft photons in
the form of $e^+e^-$ pairs about 4 to 8 times in excess of the
bremsstrahlung expectations are produced, and when hadrons are not
produced, these anomalous soft photons are also not produced
\cite{Chl84,Bot91,Bel02,Per09,DEL10}.  The transverse momenta of the
excess $e^+$$e^-$ pairs lie in the range of a few MeV/c to many tens
of MeV/c, corresponding to masses from a few MeV to many tens of MeV.
Owing to its correlated and simultaneous production alongside with
hadrons, a parent particle of an anomalous soft photon is likely to
contain elements of the hadrons, such as a pair of light quark and
light antiquark.  According to the Schwinger's $m^2$=$g^2/\pi$
relationship between the coupling constant $g$ and the boson mass $m$
of the composite fermion-antifermion pair interacting in a gauge
interaction in (1+1)D \cite{Sch62,Sch63}, the QED gauge interaction
will bring the quantized mass $m$ of a $q$$\bar q$ pair to the mass
range of the anomalous soft photons, when we consider the QCD and QED
coupling constants and the mass scale of a QCD meson.  It was
therefore proposed in \cite{Won10,Won20} that a quark and an antiquark
in a $q\bar q$ system interacting in the QED gauge interaction may
lead to new open-string bound and confined qedmesons with a mass of
many tens of MeV. These qedmesons may be produced simultaneously along
with mesons in high-energy collisions
\cite{Chl84,Bot91,Bel02,Per09,DEL10}, and the excess $e^+e^-$ pairs
may arise from the decays of the qedmesons \cite{Won10,Won20}.  The
predicted masses of the isoscalar $I(J^\pi)=0(0^-)$ and
$I(J^\pi)=1(0^-)$ qedmesons are about 13-17 and 36-38 MeV respectively
\cite{Won10,Won20}, which agree approximately with the masses of the
X17 \cite{Kra16,Kra21} and the E38 particles \cite{Abr12,Abr19}
subsequent observed.  The tentative agreement lends support to the
possible qedmeson interpretation of the anomalous particles of the
X17, the E38, and anomalous soft photons.
  
  \vspace*{-0.2cm}
\section{Can a $q$ interact with a $\bar q$ in QED alone?}

A serious question arises whether a light quark and a light antiquark
can ever be produced and interact in the QED interaction alone,
without the QCD interaction.  Actually, there are circumstances in the
decays from highly excited nuclear states with the possible production
of a light $q$$\bar q$ pair as in Fig.\ 1(a), the
hadron+hadron$\to$hadrons+$(q\bar q)^n$ reaction as in Fig.\ 1(b), and
the $e^+$+$e^-$$\to$$\gamma^*$$\to$$\,q+\bar q\,$$\to$$\,(q\bar q)^n$ reaction as in
Fig.\ 1(c), when the CM energy, $\sqrt{s}(q\bar q)$, of the produced
$q$$\bar q$ pair lies in the range $(m_q+m_{\bar
  q})$$<$$\sqrt{s}$$(q\bar q)$$<$$ m_\pi$.  In order to bring the
produced $q\bar q$ pair in this CM energy range to a possible stable
state, the produced $q$ and $\bar q$ can only interact with the QED
interaction alone, because the QCD interaction will otherwise endow
the $q$$\bar q$ pair with a CM energy beyond the range, in a
contradictory manner.

For the production of X17 in $^4$He$^*$ and $^8$Be$^*$ decays at
ATOMKI \cite{Kra16,Kra21}, we envisage the scenario that the excited
states of $^4{\rm He}(0^-\,$20.02 MeV) and $^8$Be(1$^+ \,$18.15 MeV)
are formed by pulling a proton out of one of $n$ alpha-particles of
the $(\alpha)^n$-nucleus and by placing the proton on an orbital that
is considerably outside the corresponding tritium core as shown in
Fig.\ 1(a).  The stretched string-like strong interaction between the
proton and the tritium core polarizes the vacuum so much that the
strong interaction may lead to the production of a $q$$\bar q$ pair.
At the appropriate $\sqrt{s}(q\bar q)$ eigenenergy, the QED
interaction between the $q$ and the $\bar q$ may result in the
formation of the $q$$\bar q$ bound state X17 \cite{Won10,Won20}, which
is subsequently emitted as the proton drops down to fill the hole in
the tritium core to reach the $(\alpha)^n$-nucleus ground state.  Such
a production mechanism strong suggests that X17 may also be produced
from the excited states of other $(\alpha)^n$-nuclei in which a proton
is pulled out from one of its alpha particles.  In this respect, the
$^{12}$C((1+) 18.16 MeV) state with a width of $\Gamma$=240 keV may be
an interesting analog of the $^8$Be$^*$(1$^+$ 18.15 MeV) state with a
width of $\Gamma$=138 keV and may likewise decay with the emission of
an X17 particle.

\begin{figure} [t]
\centering
\vspace*{-0.0cm}\hspace*{-0.1cm}
\resizebox{0.50\textwidth}{!}{
  \includegraphics{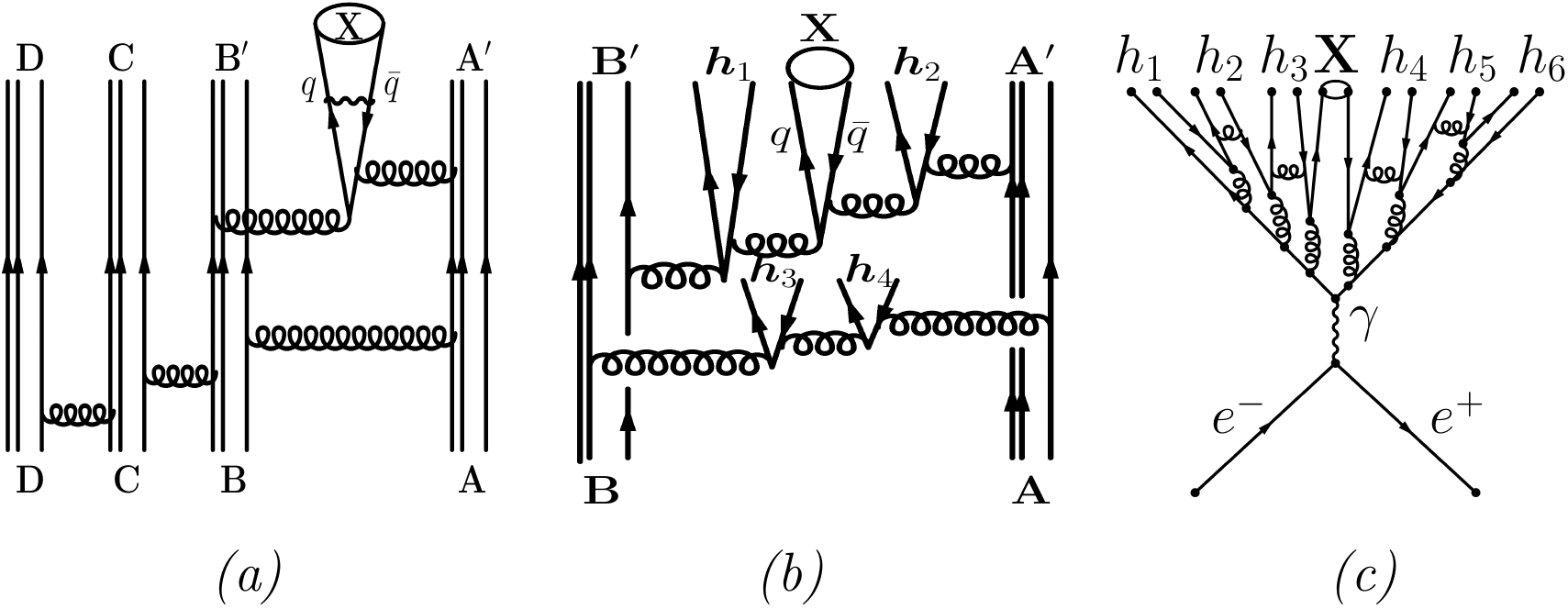}
}
\caption{Feynman diagrams for the production of a qedmeson $X$ and
  hadrons $h_i$ in (a) low-energy $^4$He$^*$ and $^8$Be$^*$ decays,
  (b) a high-energy hadron-hadron collision, and (c) a high-energy $e^+
  e^-$ annihilation.  }
\label{fig1}
\end{figure}
In other processes as illustrated in Figs.\ 1(b) and 1(c), many
$q$$\bar q$ pairs may also be produced in high-energy nuclear
collisions at Dubna \cite{Abr12,Abr19}, in high-energy hadron
collisions in anomalous soft photon production experiments
\cite{Chl84,Bot91,Bel02,Per09}, and in high-energy $e^+$-$e^-$
annihilations in DELPHI experiments \cite{Per09,DEL10}.  The $q$$\bar
q$ pairs may also be produced in high-energy heavy-ion collisions at
RHIC and LHC in two ways: either through the production of $q$$\bar q$
pairs in the multiple collision process similar to Fig.\ 1(b), or
through the coalescence of quarks in the deconfinement-to-confinement
phase transition of the quark gluon plasma.  While most produced
$q$$\bar q$ pairs will lead to hadron production, there may be
$q$$\bar q$ pairs with $(m_q+m_{\bar q})$$<$$\sqrt{s}$$(q\bar q)$$<$$
m_\pi$ for which the QED interaction between the quark and the
antiquark may lead to the production of the X17 and E38 particles at
the appropriate energies.  The observation of the E38 particle at
Dubna suggests that along with E38, the X17 particle with a mass of 17
MeV may also be produced in the same reaction at Dubna with an even
greater probability because of its lower mass.

\section{How\,can\,$q\bar q$\,be\,produced\,and\,confined\,in\,QED?}

To answer the question how a $q\bar q$ pair can be produced and
confined in a QED meson, we note first of all that there are two
different types of QED U(1) gauge interactions possessing different
confinement properties \cite{Pol77,Pol87,Dre79}.  There is the compact
QED U(1) gauge theory in which the gauge fields $A^\nu$ are angular
variables with a periodic gauge field action that allows transverse
photons to self-interact among themselves.  Defined on a
lattice, the compact QED U(1) gauge theory has the gauge field action
\cite{Pol77,Pol87,Dre79}
\begin{eqnarray}
S=\frac{1}{2g^2}\sum_{x,\alpha \beta} (1-\cos F_{x,\alpha \beta}),
\end{eqnarray}
where $g$ is the coupling constant and the gauge fields are 
\begin{eqnarray}
F_{x, \alpha \beta}\!=\!A_{x,\alpha}\!\!+\!\! A_{x+\alpha,\beta}\!\! - \!\! A_{x+\beta,\alpha} \!\! - \!\! A_{x,\beta},{\rm with} -\!\!  \pi\!\!  \le A_{x,\alpha} \!\! \le\!\!  \pi.~
\label{eq2}
\end{eqnarray}
There is also the non-compact  QED U(1) gauge theory with the gauge
field action \cite{Pol77,Pol87,Dre79}
\vspace*{-0.1cm}
\begin{eqnarray}
S=\frac{1}{2g^2}\sum_{x,\alpha \beta} F_{x,\alpha \beta}^2,  ~~~{\rm with}~~~~- \infty  \le A_{x,\alpha} \le  +\infty  .
\end{eqnarray}
\vspace*{-0.0cm}
In non-compact QED gauge theories, the transverse photons do not
interact with other transverse photons.  Even though the compact and
the non-compact QED gauge theories have the same continuum limit, they
have different confinement properties.  A pair of opposite electric
charges are confined in compact QED in (2+1)D and strong coupling
(3+1)D, but they are unconfined in weak coupling (3+1)D
\cite{Pol77,Pol87,Dre79}.  They are unconfined in non-compact QED in
(3+1)D \cite{Pol77,Pol87,Dre79}.

Which type of the QED U(1) gauge interaction does a quark interact
with an antiquark?  As pointed out by Yang \cite{Yan70}, the
quantization and the commensurate properties of the electric charges
of the interacting particles imply the compact property of the
underlying QED gauge theory.  Because (i) quark and antiquark electric
charges are quantized and commensurate, (ii) quarks and antiquarks are
confined, and (iii) there are experimental evidences for possible
occurrence of confined $q\bar q$ qedmeson states as we mentioned in
the Introduction, it is therefore reasonable to propose that quarks
and antiquarks interact with the compact QED U(1) interaction.

In compact QED, Polyakov \cite{Pol77,Pol87} showed previously that a
pair of opposite electric charges in (2+1)D$_{\{x^1,x^2,x^0\}}$
space-time are confined, and that the confinement persists for all
non-vanishing coupling constants, no matter how weak.  As explained by
Drell and collaborators \cite{Dre79}, such a confinement in
(2+1)D${}_{\{x^1,x^2,x^0\}}$ arises from the angular-variable property
of $A_\phi$ and the periodicity of the gauge field action as indicated
in Eqs.\ (1) and (2).  Such gauge periodicity in the neighborhood of
the produced opposite electric charges leads to self-interacting
transverse gauge photons.  These transverse gauge photons interact
 among themselves, they do not radiate away, and they join the
two opposite electric charges by a confining linear interaction.

\begin{figure} [h]

\centering
\vspace*{-0.5cm}\hspace*{-0.3cm}
\resizebox{0.44\textwidth}{!}{
\includegraphics{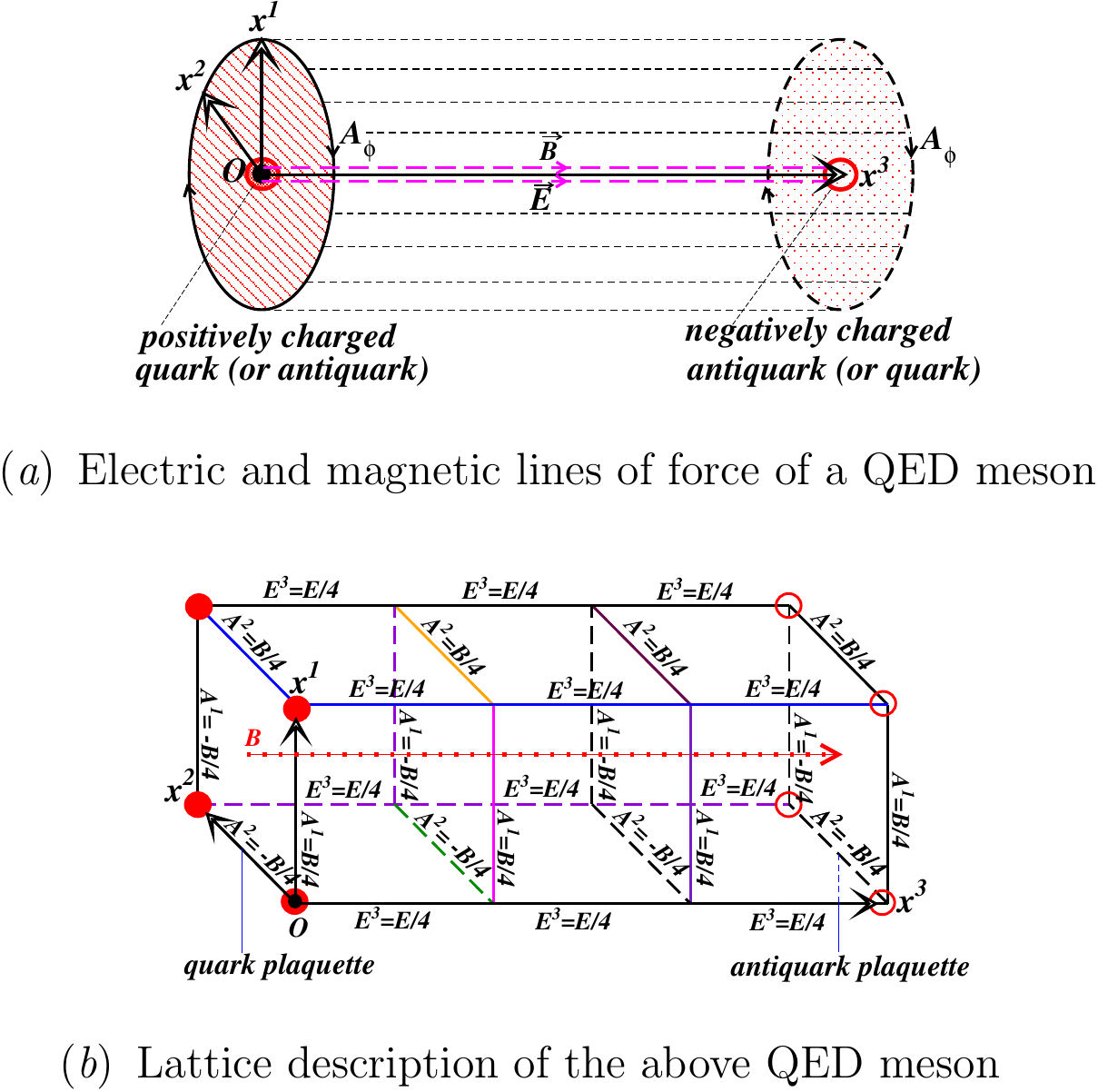}}
\caption{}
\vspace*{-0.3cm}
\end{figure}
\vspace*{-0.1cm} We can use the above Polyakov's result in compact QED
in (2+1)D as the starting point to construct a model of a quark and an
antiquark produced and confined in a QED meson.  We envisage the
production of the $q$$\bar q$ pair at the eigenenergy $\sqrt{s}(q\bar
q)$ of the QED meson and take the quark charge to be positive, which
can be easily generalized to other cases.  We consider the production
of the $q\bar q$ charge pair at $(x^1,x^2,x^3)$=0 with the antiquark
separated initially from the antiquark along a direction chosen to be
the longitudinal $x^3$ direction at an incipient separation $\Delta
x^3$.  The creation of the $q$$\bar q$ charge pair is accompanied by
the creation of the gauge fields $\bb A$, $\bb E$, and $\bb
B$(=$\nabla $$\times $${\bb A})$.  We can apply Polyakov's result to
infer that the produced charges and the QED gauge fields are confined
in (2+1)D$_{\{x^1,x^2,x^0\}}$ transversely  at the $x^3$$\sim 0$ plane.  We can now
stretch the antiquark longitudinally along $x^3$ away from the quark
at $x^3$=0 to execute the 3D yo-yo motion for the bound state,
reaching a momentary snapshot of the flux tube in the stretch (2+1)D
configuration shown in Fig.\ 2(a).  We can transcribe Fig.\ 2(a)  in
terms of the lattice link and plaquette variables in Fig.\ 2(b), by
following the Hamiltonian formulation and the notations of Drell
$et~al.$ \cite{Dre79}.  Specifically, in the $A^0$=0 gauge we specify
the canonical conjugate gauge fields $\bb A$ and $\bb E$ at the links
in Fig.\ 2(b), where we display only the $A^1,A^2$ and $E^3$ values of the conjugate gauge fields.  The magnetic field $\bb B$ associated with the
plaquettes can be determined as the curl of $\bb A$ and is directed
along $x^3$ in Figs.\ 2(a) and 2(b).  The magnetic field $\bb B$ sends
the quark and antiquark charges into the appropriate Landau orbitals
to execute transverse zero-mode  harmonic oscillator zero-point motions on
their $\{x^1,x^2\}$ planes.  At the qedmeson eigenenegy, the electric
field $\bb E$ and the magnetic field $\bb B$ along the longitudinal
$x^3$ direction send the quark and the antiquark in longitudinal 3D
yo-yo motion.  The positive electric quark charge fractions (solid
circles in Fig.\ 2(b)) reside at the $x^3$=0 plaquette vertices and
the negative electric antiquark charge fractions (open circles in
Fig.\ 2(b)) at the antiquark plaquette vertices at the $x^3$ plane.
The transverse gauge fields $\bb A$ on the transverse links are copies
of those on the quark and the antiquark plaquettes, and they are
unchanged in the stretching, while the longitudinal links are all
$E^3$=$|\bb E|/4$. Consequently, the self-interactions of the transverse gauge
fields that initially confine the charges and the gauge field
transversely will be retained and the cloud of gauge fields will
continue to interact with each other to maintain the
transverse confinement.

With transverse confinement and $\bb E$ and $\bb B$ aligned along the
$x^3$ direction, it remains necessary to study longitudinal
confinement, dynamical quark effects, and spontaneous chiral symmetry
breaking.  Therefore, we 
idealize the flux tube in stretch (2+1)D as a longitudinal string in
(1+1)D$_{\{x^3,x^0\}}$ and
approximate the quarks to be massless.  With massless quarks in QED in (1+1)D$_{\{x^3,x^0\}}$, there is
a gauge-invariant relation between the quark current
$j^\mu$ and the gauge field $A^\mu$ as given by
\cite{Sch62,Sch63,Won94}
\vspace*{-0.2cm}
\begin{eqnarray}
j^\mu = -\frac{g}{{\pi}} ( A^\mu 
- \partial^\mu \frac{1}{\partial_\lambda \partial^\lambda} \partial_\nu A^\nu),
~~~\mu,\lambda,\nu =0,3. 
\end{eqnarray}
On the other hand, the gauge field $A^\nu$ depends on the quark
current $j^\nu$ through the Maxwell equation,
\begin{eqnarray}
\partial_\nu(\partial^\mu A^\nu- \partial^\nu A^\mu)=-g j^\mu,~~~\mu,\nu =0,3.
\end{eqnarray}
Equations (4) and (5) lead to $- \Box A^\mu$=$(g^2/\pi)A^\mu$ 
and\break  $- \Box j^\mu$=$(g^2/\pi)j^\mu$, with  $j^\nu$ and
 $A^\nu$ self-interacting among themselves 
and building a longitudinal confining interaction between the
quark and the antiquark in (1+1)D.  As a consequence, in accordance with
Schwinger's exact solution for massless fermions in QED in (1+1)D
\cite{Sch62,Sch63}, the light quark and the light antiquark
interacting in QED will be longitudinally confined just as well and
will form a stable QED quark-antiquark system.  Possessing both
transverse and longitudinal confinements as in an open-string, such a
stable QED state may be experimentally observed as a qedmeson.  By
using the method of bonsonization, we obtain the masses of the
lowest-energy states of the open-string QED mesons which adequately
match those of X17 and E38 \cite{Won10,Won20} to support its
approximate validity.

\section{How can  QED mesons be detected? }

The qedmesons can be detected by the invariant masses of their decay
products.  In a qedmeson, the quark and the antiquark  can annihilate,
leading to the emission of two real photons ($\gamma_1\gamma_2)$ as in
Fig.\ 3(a), two virtual photons ($\gamma_1^*\gamma_2^*)$ or two
dilepton $(e^+e^-)$ pairs as in Fig.\ 3(b), or a single $(e^+e^-)$
pair as in Fig.\ 3(c).  \setcounter{figure}{2}
\begin{figure}[h]
\centering
\resizebox{0.48\textwidth}{!}{
\includegraphics{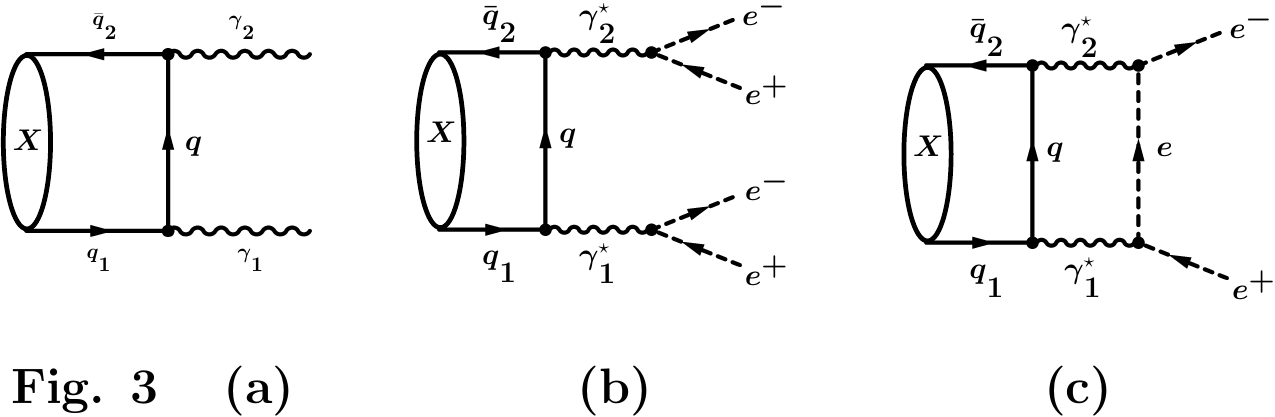}}
\vspace*{-0.2cm}
\label{fig3}
\end{figure}
\vspace*{-0.0cm}
We can make an order of magnitude estimate on the decay width of X17
into two photons as depicted in Fig.\ 3(a).  
From
Eq.\ (89.3) of \cite{Ber79}, we have for X17$\to \gamma \gamma$,
\vspace*{-0.1cm}
\begin{eqnarray}
\Gamma( {\rm X17}\! \to\! \gamma \gamma)&&\!\!=\!\! 
\bigg \{ \frac{1}{2}\bigg [ \bigg (\frac{1}{3} \bigg )^2 \!\!+\! \bigg ( \frac{2}{3}\bigg  )^2 \bigg ] \bigg \}^2 
\frac{4 \pi  \alpha_{\rm QED}^2}{M^2} |\psi(0)|^2,\hspace*{0.4cm}
\end{eqnarray}
where $\psi(0)$ is the wave function at the origin.  The wave function
at the origin can be estimated from the size of the qedmeson, $
|\psi(0)|^2$$ \sim $$1/{\pi R_T^2 L_z}$, where $R_T$$\sim$0.4 fm
\cite{Won20} and the longitudinal length $L_z$$\sim$7.15 fm as
estimated from Table 2 for the lowest qedmeson state in
Ref.\ \cite{Won20a}.  For the X17 with a mass of 17 MeV, $\Gamma({\rm
  X17}\to \gamma \gamma) \sim 0.4$ MeV.

From the total width and the branching ratio into $e^+e^-$ in 
ATOMKI measurements \cite{Kra16,Kra21}, the X17 width $\Gamma({\rm
  X17} $$\to$$ ~e^+ e^-)$ can be estimated to be 4.2 eV from $^4$He
decay and 0.828 eV from $^8$Be decay.  The knowledge of the approximate
widths may facilitate future searches for the qedmesons.

In high-energy heavy-ion collisions at RHIC and LHC, one expect
copious production of $q\bar q$ pairs either from the multiple
collision process or from the coalescence of quarks and antiquarks in
the confinement-to-deconfinement phase transitions.  Among the
produced $q \bar q$ pairs, there will be some pairs whose invariant
masses match the qedmeson eigenenergies to lead to the production of
qedmesons.  For these produced qedmesons, the decay into two virtual
photons via Fig.\ 3(b) offers an interesting tool for the search of
the anomalous particles.  Specifically, a decay into two virtual
photons involves the measurement of the momenta of four final leptons
which requires a high degree of correlation.  As a consequence, the
experimental noises of chance coincidences may be significantly
reduced.  One can construct the sum and the difference of the
invariant momenta square of the virtual photon 4-momenta, $P$$=$$
\sqrt{(p_{\gamma_1^*}+p_{\gamma_2^*})^2}$, and $Q$=$\sqrt{-
  (p_{\gamma_1^*}-p_{\gamma_2^*})^2}$.  The virtual diphoton pair
distribution ${dN(P,Q)}/{dP~dQ}$ at RHIC and LHC will provides useful
information to search for the qedmesons.

\vspace*{-0.3cm}
\section{Conclusions and Discussions}
\vspace*{-0.0cm} The observations of the X17 particle, the E38
particle, and the anomalous soft photons raise many interesting
questions on the nature of these anomalous particles. While most
theoretical discussions center on the elementary particle
possibilities \cite{X1722}, we examine here the description of these
anomalous particles as composite states of a quark and an antiquark
interacting in compact QED.  We propose a model of the production and
the confinement of a quark and an antiquark in a QED meson by
combining Polyakov's transverse confinement of opposite electric
charges in compact QED in (2+1)D and Schwinger's longitudinal
confinement in QED in (1+1)D.  The important ingredients are (i) the
self-interactions of the transverse photons in compact QED that
confine the gauge fields and the opposite electric charges
transversely in (2+1)D\cite{Pol77,Pol87}, (ii) the stretching of the
(2+1)D with incipient longitudinal separation to come to the
longitudinal flux tube configuration in a 3D yo-yo motion, (iii) the
idealization of the longitudinal flux tube of the stretch (2+1)D
configuration as a string in (1+1)D and the light quarks as massless,
and finally (iv) Schwinger's solution of longitudinal confinement of
massless quarks in QED (1+1)D \cite{Sch62,Sch63}, resulting in a bound
and confined $q\bar q$ as a composite qedmeson state.  The quark and
the antiquark in a qedmeson are essentially electric charge monopoles
and Polyakov's magnetic monopoles in dynamical motion.  If this
picture of a $q$ and a $\bar q $ interacting in compact QED
interactions is correct, it will imply that a quark and an antiquark
obey QED laws that differ from those for an electron and a positron.

It will be of great interest to study in future lattice gauge
calculations the problem of quark confinement and $q\bar q$ bound
states in compact QED for quarks with different color and flavors in
the stretch (2+1)D configuration, with the proper quantization and
dynamical light quarks.

 \end{document}